\documentclass[aps,pra,twocolumn,showpacs]{revtex4}
\usepackage[pdftex]{graphicx}
\usepackage{amssymb}
\usepackage{amsmath}
\usepackage{bm}

\DeclareGraphicsExtensions{.pdf}

\begin{document}

\title{Critical velocity for vortex shedding in a Bose-Einstein condensate}
\author{Woo Jin Kwon, Geol Moon, Sang Won Seo, and Yong-il. Shin}

\email{yishin@snu.ac.kr}

\affiliation{Department of Physics and Astronomy, and Institute of Applied Physics, Seoul National University, Seoul 151-747, Korea}

\begin{abstract}
We present measurements of the critical velocity for vortex shedding in a highly oblate Bose-Einstein condensate with a moving repulsive Gaussian laser beam. As a function of the barrier height $V_0$, the critical velocity $v_c$ shows a dip structure having a minimum at $V_0 \approx \mu $, where $\mu$ is the chemical potential of the condensate. At fixed $V_0\approx 7\mu$, we observe that the ratio of $v_c$ to the speed of sound $c_s$ monotonically increases for decreasing $\sigma/\xi$, where $\sigma$ is the beam width and $\xi$ is the condensate healing length.  We explain our results with the density reduction effect of the soft boundary of the Gaussian obstacle, based on the local Landau criterion for superfluidity. The measured value of $v_c/c_s$ with our stiffest obstacle is about 0.4, which is in good agreement with theoretical predictions for a two-dimensional superflow past a circular cylinder.
\end{abstract}


\maketitle

\section{Introduction}

A superfluid flows without friction but becomes dissipative above a certain critical velocity $v_c$ via generating its elementary excitations such as phonons and vortices. The Landau criterion provides a conventional energetic consideration to determine the critical velocity, stating $v_c=\min[\epsilon(p)/p]$~\cite{Landau}, where $\epsilon(p)$ is the energy of an elementary excitation of momentum $p$. For a homogeneous system, the Landau critical velocity is equal to the speed of sound $c_s$. However, the dynamic response of a superfluid flow is significantly sensitive to the boundary condition of the system and hence a quantitative understanding of the critical velocity has been a challenging task in the study of superfluidity.

One of the paradigmatic situations considered in fluid mechanics is a two-dimensional (2D) flow past a circular cylinder. For an incompressible flow, the local velocity is increased by a factor of 2 at the lateral sides of the cylinder~\cite{Landau2} and the local Landau supersonic criterion suggests a critical velocity of $v_c=0.5c_s$ that is independent of the radius $R$ of the cylinder. Theoretical studies showed that the onset of dissipation involves generating a counter-rotating vortex pair~\cite{frisch,jackson1}. More rigorous calculations, taking into account the compressibility of the superfluid and quantum pressure near the boundary of the cylinder, predicted that the critical velocity converges to $v_c= 0.37c_s$ in the large cylinder limit $R\gg\xi$~\cite{berloff2,brachet1,rica,brachet3,berloff}, where $\xi$ is the superfluid healing length. Experimental verification of the predictions on $v_c/c_s$ is highly desirable.

In previous ultracold atom experiments, a similar situation was investigated by stirring superfluid samples with a repulsive laser beam~\cite{raman,onofrio,inouye,neely,dalibard,kwon}. The existence of finite critical velocities~\cite{raman,onofrio,dalibard} and generation of vortex dipoles~\cite{inouye,neely,kwon} were successfully demonstrated. However, the measured values of $v_c/c_s$ ranged widely from 0.1 to 0.45, which did not allow a quantitative study of the homogeneous 2D problem. Theoretical investigations showed that  the inhomogeneous density distribution of a trapped sample~\cite{winiecki,crescimanno,fedichev}, 3D vortex dynamics~\cite{brachet2,pomeau,jackson3}, or the manner of stirring~\cite{winiecki,pomeau,tsubota,neely,stagg2} should be critical in the measurements. Recently, the vortex shedding dynamics was also investigated experimentally with polariton superfluids flowing past static defects~\cite{nardin,amo}.

In this paper, we systematically study the critical velocity for vortex shedding in a Bose-Einstein condensate with a repulsive Gaussian potential. We measure the critical velocity as a function of the barrier height $V_0$ of the potential over a wide range of the beam width, $10<\sigma/\xi<55$. In particular, in order to address the 2D homogeneous regime, we employ spatially large and highly oblate condensates, ensuring vortex dynamics in two dimensions.

The key difference of a Gaussian potential from a hard cylinder is its soft boundary. A Gaussian potential, $V(r) = V_{0} \exp(-2r^{2}/\sigma^{2})$, produces a density-depleted hole in the condensate when $V_0>\mu$, where $\mu$ is the chemical potential of the condensate. The radius of the hole and the potential slope at the hole boundary are given as 
\begin{eqnarray}
R&=&\sigma \sqrt{\ln(V_{0}/\mu)/2}\\
S&=&-\frac{dV}{dr}\Big|_{r=R}=4\mu R/\sigma^2,
\end{eqnarray}
respectively. In comparison to the case with a hard cylinder, the soft boundary reduces the density in the proximity of the obstacle and consequently lowers the local speed of sound. Then, from the local Landau criterion it is naturally suggested that the critical velocity of the hard cylinder defines an upper bound for that of the obstacle formed by the Gaussian potential. When the hole radius $R$ becomes larger with higher $V_0$ [Fig.~1(b)] and/or the beam width $\sigma$ decreases for fixed $R$ [Fig.~1(c)], the obstacle would converge to the hard cylinder with stiffening its boundary. 

The main result of our measurements is that in the deep non-penetrable regime (i.e., $V_0\gg\mu$), the critical velocity $v_c$ increases with decreasing $\sigma/\xi$ for fixed $V_0/\mu$ and approaches about $0.4c_s$. This observation is consistent with the expectation from the aforementioned discussion based on the local Landau criterion. Furthermore, the measured value of $v_c/c_s$ with our stiffest obstacle is in good agreement with theoretical predictions for a 2D superfluid flow past a hard cylinder.

\begin{figure}
\includegraphics[width=8.7cm]{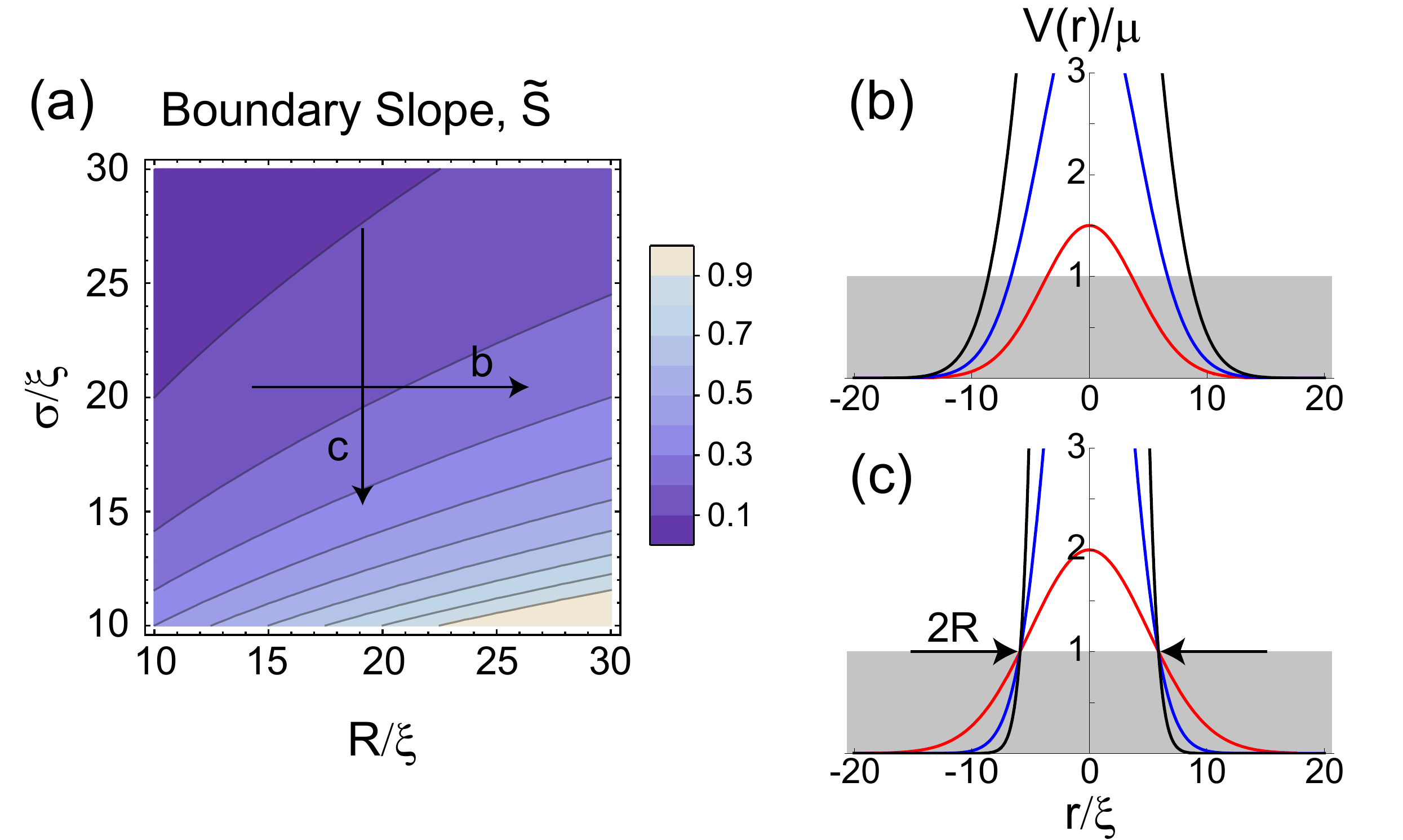}
\caption{(color online). Soft boundary of the optical obstacle formed by a repulsive Gaussian potential $V(r)=V_0 \exp(-2r^2/\sigma^2)$. (a) Normalized potential slope at the obstacle boundary $\tilde{S}=(\xi/\mu)|dV/dr|_{r=R}$ as a function of $\sigma/\xi$ and $R/\xi$, where $\mu$ and $\xi$ is the chemical potential and the healing length of the condensate, respectively, and $R$ is the obstacle radius such that $V(R)=\mu$. The boundary becomes stiffer as (b) $R/\xi$ becomes larger or (c) $\sigma/\xi$ decreases. The corresponding trajectories are indicated by the arrows in (a).}
\label{Fig1new}
\end{figure}

\section{Experiment}

Our experiment starts with a Bose-Einstein condensate of $^{23}$Na atoms in a harmonic trap formed by combining optical and magnetic potentials~\cite{kwon}. The condensate fraction of the sample is over 90\%. In a typical sample condition, where the trapping frequencies are $\omega_{r,z}$ = $2 \pi \times (9.0, 400)$ Hz and the atom number of the condensate is $N_{0} = 3.2(2) \times 10^{6}$, the condensate healing length is $\xi=\hbar/\sqrt{2m\mu}\approx 0.46~\mu$m and the speed of sound is $c_s=\sqrt{\mu/m}\approx 4.3~$mm/s at the trap center, where $\hbar$ is the Planck constant divided by 2$\pi$ and $m$ is the atomic mass. By adjusting the trapping frequencies or the atom number of the condensate, $\xi$ is varied up to $0.9~\mu$m. The Thomas-Fermi radius and thickness of the condensate are $R_\mathrm{TF}/\xi\geq240$ and $Z_\mathrm{TF}/\xi\leq6$, respectively. In this highly oblate condensate, vortex line excitations are strongly suppressed~\cite{zaremba,rooney} and the vortex dynamics is expected to be two dimensional. 

We adiabatically ramp up the power of a repulsive Gaussian laser beam in 1~s and hold it for 0.2~s to ensure that the condensate is stationary. Then, we translate the laser beam horizontally by $24~\mu$m by using a piezo-driven mirror [Fig.~2(a)]. The velocity $v$ of the laser beam is kept constant during the translation and controlled by adjusting the traveling time. The sweeping region of the laser beam is centered in the condensate [Fig~2(b)]. The density variation over the sweeping region is less than 10\% and the speed of sound can be well approximated to be spatially constant. After completing the sweeping, we slowly ramp down the laser beam power for 0.5~s, and take an absorption image of the condensate after expansion by releasing the trapping potential to detect vortices.

\begin{figure} 
\includegraphics[width=7.8cm]{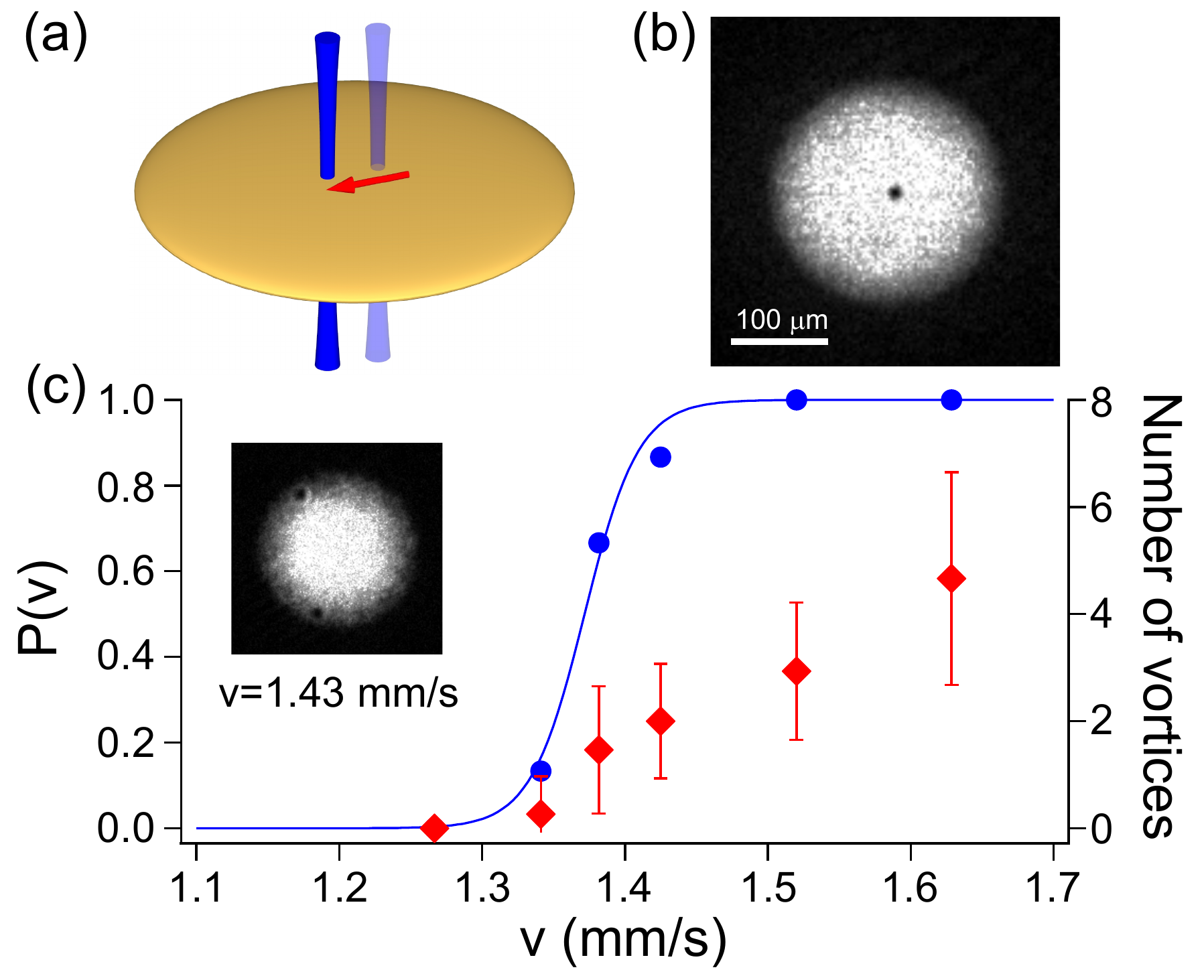}
\caption{(color online). Vortex shedding in a highly oblate Bose-Einstein condensate. (a) Schematic of the experiment. An optical obstacle is formed by a repulsive Gaussian laser beam penetrating through the condensate and moves horizontally at a constant velocity in the center region. (b) An in-situ image of the condensate with the optical obstacle at the initial position. (c) Probability $P(v)$ of having vortex dipoles (blue circles) and the number of vortices (red diamonds) as a function of the velocity $v$ of the optical obstacle. The barrier height $V_{0}\approx2.1 \mu$ and the Gaussian width $\sigma\approx 20\xi$.  The solid line is a sigmoidal function fit to $P(v)$. The inset shows an image of a condensate containing a vortex dipole. Each data point was obtained from 15 realizations of the same experiment and the error bars indicate the standard deviations of the measurements.}
\label{Fig2}
\end{figure}

The width $\sigma$ of the laser beam is calibrated from the in-situ images of very dilute samples penetrated through by the laser beam, taking into account the imaging resolution of our system ($\approx 5~\mu$m). The beam waist of the laser beam is $9.1(12)~\mu$m at the focal plane and the Gaussian width $\sigma$ of the optical obstacle is controlled by defocusing the laser beam at the sample plane. The sample thickness ($<3~\mu$m) is much shorter than the Rayleigh length of the laser beam and we ignore the beam divergence. The beam width is much smaller than the condensate radius and the chemical potential of the condensate is negligibly affected by the presence of the laser beam.

A vortex dipole is identified with two density-depleted holes that are symmetrically located in the condensate with respect to the sweeping line of the laser beam~[Fig.~2(c) inset]. After beinggenerated in the center region of the condensate, the vortex dipole moves toward the edge of the condensate and splits into two individual vortices that subsequently travel along the boundary of the condensate in the opposite direction. This peculiar obrit motion of the vortices was studied in Ref.~\cite{neely}. A single vortex was observed occasionally at low $v$ with probability $<4 \%$. Because the vortex lifetime is over 10~s~\cite{kwon}, we attribute the single vortex to uncontrolled perturbations in sample preparation and we do not count it as a vortex dipole.

The critical velocity $v_c$ for vortex shedding is determined from the probability distribution $P(v)$ for having vortex dipoles after sweeping with the laser beam. Here $P(v)$ is obtained from 15 realizations of the same experiment with a given sweeping velocity $v$, i.e., $P(v)$ is the ratio of the number of images showing vortex dipoles to the total number of measurements. The critical velocity $v_c$ is determined by fitting a sigmoidal function to the probability distribution as $P(v)=1/(1+e^{-(v-v_{c})/\gamma})$ [Fig.~2(c)]. We use the value of $1.5\gamma$ as the measurement uncertainty of $v_{c}$, corresponding to the range of $0.2\leq P \leq 0.8$.

\section{Results and Discussion}

Figure 3(a) displays the results of the critical velocity as a function of the barrier height $V_0$ for various beam widths. The critical velocity shows a dip structure having a minimum at $V_0\approx \mu$, clearly distinguishing the two regimes: a penetrable regime with $V_0<\mu$ and a non-penetrable regime with $V_0>\mu$. The dip structure of the critical velocity can be accounted for by a consideration based on the local Landau criterion. For $V_0<\mu$, the density minimum is located at the top of the Gaussian potential and decreases for higher $V_0$, and thus, lowering the local speed of sound. On the other hand, when $V_0>\mu$, as described before, the potential slope becomes steeper with higher $V_0$ and the density in the proximity of the obstacle boundary is gradually restored back to the bulk density, leading to a higher critical velocity. The dip structure around $V_0=\mu$ becomes more pronounced with larger $\sigma/\xi$ by lowering the minimum value [Fig.~3(a) inset].

\begin{figure}
\includegraphics[width=8.3cm]{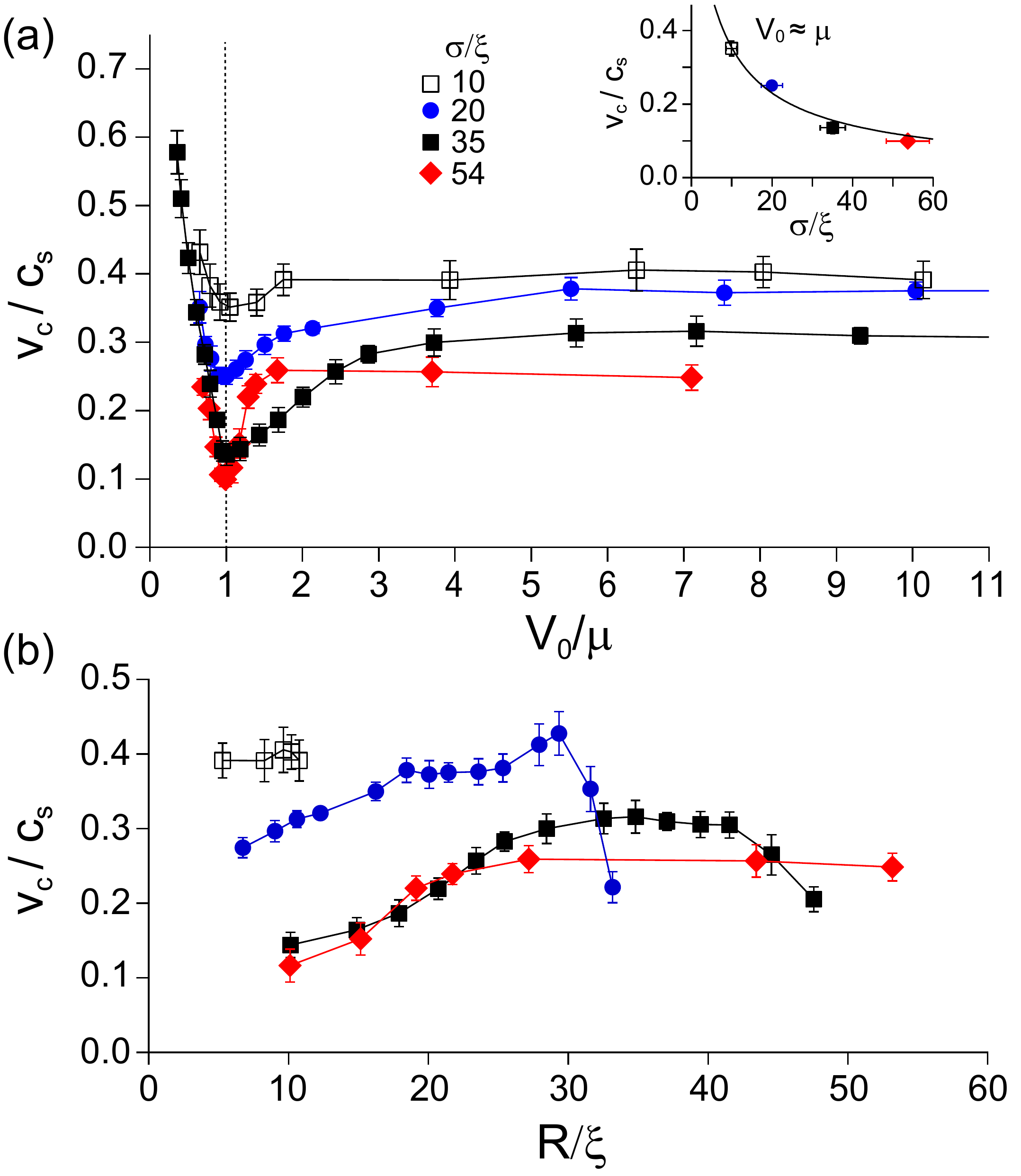}
\caption{(color online). (a) Normalized critical velocity $v_{c}/c_s$ versus the relative barrier height $V_{0}/\mu$ for various beam widths $\sigma/\xi$: $(\sigma,\xi)=[24.5(24),0.46]$ (red diamonds), $[16.0(14),0.46]$ (black squares), $[9.1(12),0.46]$ (blue circles), and $[9.1(12),0.9]~\mu$m (open squares). The inset displays $v_{c}/c_s$ at $V_{0} \approx \mu$ as a function of $\sigma/\xi$ and the solid line is a line to guide the eye. (b) Same data in the non-penetrable regime ($V_0>\mu$) as a function of the hole radius $R/\xi$, together with additional data points for $V_0/\mu>10$.}
\label{Fig3}
\end{figure}

In terms of the vortex shedding mechanism, the penetrable regime is different because there is no density-depleted region in the fluid and it has been anticipated that vortex nucleation would be initiated by generation of rarefaction pulses~\cite{brachet3,berloff,saito,jackson2}. Recently, the critical velocity of penetrable obstacles for dissipation was investigated in effective 1D systems~\cite{engels,leboeuf,hulet,campbell}.  In one dimension, the critical velocity is predicted to vanish as $V_0$ approaches $\mu$~\cite{leboeuf} but it is not the case in our 2D situation. Moreover, the critical velocity shows a quite intriguing dependence on $\sigma/\xi$ [Fig.~3(a) inset]. Further investigation of the functional form of $v_c(V_0/\mu;\sigma/\xi)$ in the penetrable regime is warranted. However, in this work, we focus on the non-penetrable regime to address the 2D hard cylinder situation.

In Fig.~3(b), we recast the data for the non-penetrable regime ($V_0>\mu$) as a function of the hole radius $R$, together with additional data obtained for $V_0>10\mu$ with $\sigma=20\xi$ and $35\xi$. Note that the hole radius is weakly dependent on the barrier height as $R\propto \sqrt{\ln (V_0/\mu)}$. From the previous discussion of the boundary stiffness effect and $S\propto R/\sigma^2$~[Fig.~1(b)], one may expect that $v_c/c_s$ would be saturated to a certain value with increasing hole radius and also that the saturation behavior would be faster with smaller beam width. We see that the experiment data are roughly fit to the expectation in a small $R$ region. However, when $R$ is increased to be larger than $\sigma$, the growth rate of the critical velocity slows down and becomes negative. Note that at $R/\xi\sim 35$, the critical velocity with small $\sigma/\xi=20$ is even lower than that with large $\sigma/\xi=54$.

One possible explanation for such climbing-over behavior of the critical velocity is the imperfection of the laser beam profile. If the beam profile is not perfectly Gaussian, for example, the intensity profile of the outer part of the laser beam decays slower than exponential and then the potential slope $S$ at the obstacle boundary would decrease with increasing $R$. We see that the climbing-over of $v_c$ occurs at $R\sim \sigma$ in both of the data sets with $\sigma/\xi=20$ and 35. This seems to support the beam profile effect because in our experiment $\sigma$ is varied by defocusing the same laser beam. The $M^2$ factor of the laser beam is measured to be 1.2.

To further investigate the soft boundary effect on the critical velocity, we take a different scanning trajectory in the parameter space of the Gaussian obstacle: decreasing $\sigma/\xi$ for fixed $V_0/\mu$. In this setting, the hole radius $R$ is also varied proportionally with $\sigma$ [Eq.~(1)] and the trajectory corresponds to a diagonal line in Fig.~1(a). A scanning with fixed $R/\xi$, as depicted in Fig.~1(c), might be more ideal in terms of isolating the finite-$R/\xi$ effect~\cite{brachet1,brachet3,berloff}, but this would require exponentially high $V_0/\mu$ for small $\sigma/\xi$ as $V_0/\mu = e^{2(R/\sigma)^2}$, necessarily recalling the outer part of the laser beam. We set $V_0/\mu\approx 7$, where $R\approx \sigma$ and the normalized boundary slope $\tilde{S}=(\xi/\mu)S\approx 4(\sigma/\xi)^{-1}$. In the previous measurements (Fig.~3), $v_c(V_0/\mu)$ shows a maximum around this potential height.  

In the new set of measurements, we observe that the critical velocity monotonically increases as  $\sigma/\xi$ decreases (Fig.~4), which is consistent with our expectation from the boundary stiffness. It is worth noting that although the data are obtained from various samples with different healing lengths, they agree with each other in the plane of the dimensionless parameters $\sigma/\xi$ and $v_c/c_s$. This demonstrates the 2D character of the vortex dynamics in our system because the healing length is the only relevant length scale in 2D superfluid hydrodynamics.

\begin{figure}
\includegraphics[width=8cm]{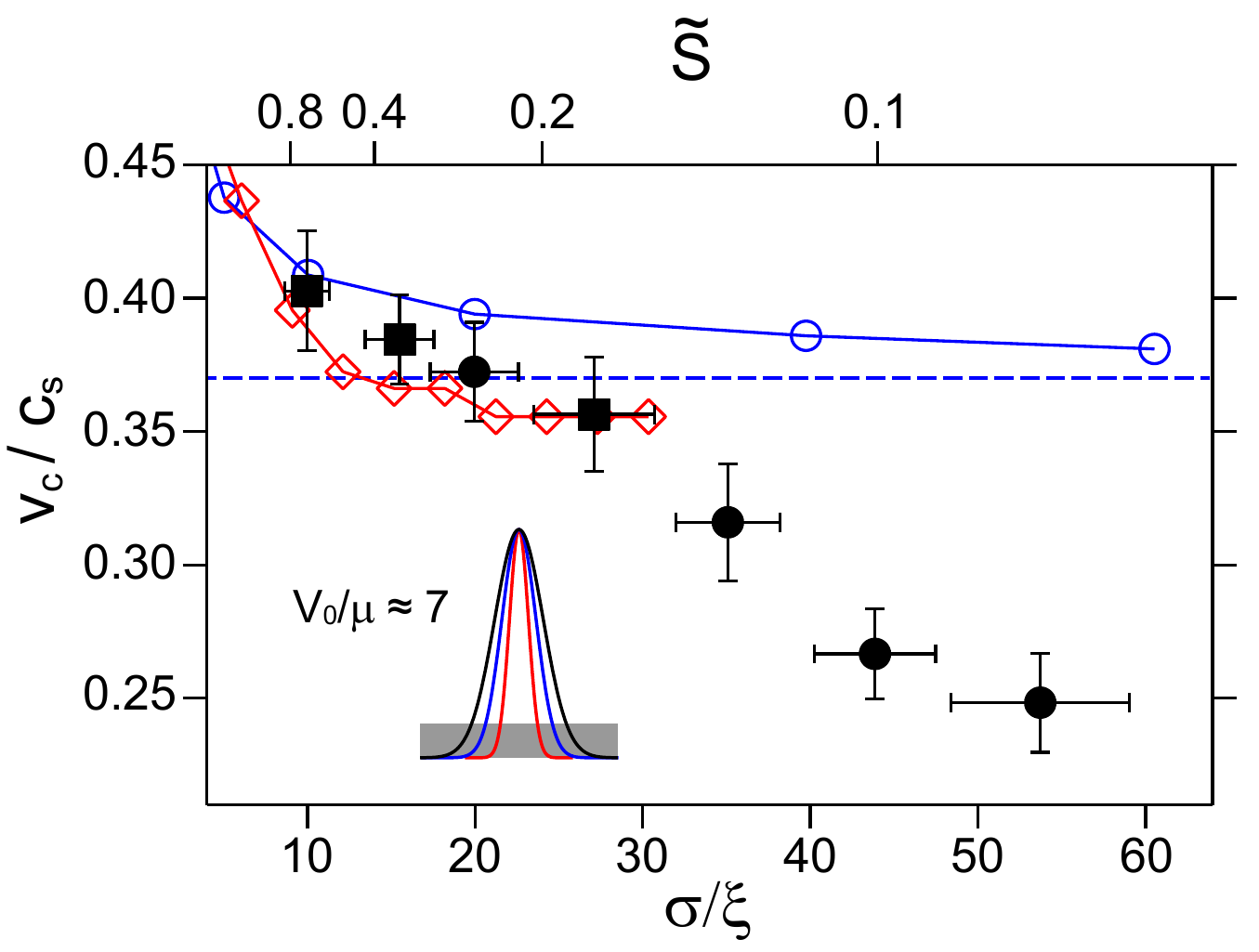}
\caption{(color online). Normalized critical velocity $v_{c}/c_s$ versus $\sigma/\xi$ for fixed $V_{0}/\mu \approx 7$. Here $\sigma=9.1(12)~\mu$m for black closed squares and $\xi=0.46~\mu$m for black closed circles. The blue open circles show the results of theoretical calculation for a 2D hard cylinder of radius $R=\sigma\sqrt{\ln{(V_0/\mu)}/2}$ from Ref.~\cite{brachet3}. The red open diamonds indicate the numerical results for a Gaussian potential with $V_0/\mu=100$ from Ref.~\cite{stagg}, where the potential slope $\tilde{S}$ would be about two times higher than ours. The dashed line denotes the theoretically predicted value, $v_{c}/c_s=0.37$ for a 2D homogeneous case with a hard cylinder in the large obstacle limit~\cite{rica,brachet3,berloff}.} 
\label{Fig4}
\end{figure}

For our stiffest obstacle, $v_c/c_s$ is measured to be about 0.40, which is slightly higher than the predicted value of 0.37 for a 2D hard cylinder in the large-$R/\xi$ limit. In our measurement, $R\geq 10\xi$ marginally satisfies the large obstacle condition, and the deviation of the measured value might be attributed to the finite-$R/\xi$ effect. The dependence of the critical velocity on $R/\xi$ was investigated theoretically~\cite{brachet1,brachet3,berloff} and it was shown that $v_c/c_s$ gradually increases from the value of 0.37 as $R/\xi\rightarrow 0$. In Fig.~4, for comparison, we display the theoretical results of Ref.~\cite{brachet3} for a 2D hard cylinder of radius $R=\sigma\sqrt{\ln(V_0/\mu)/2}$ (blue open circles) and the result of numerical simulations performed with a Gaussian potential with $V_0/\mu=100$ in Ref.~\cite{stagg} (red open diamonds). The experimental results converge to the theoretical predictions when $\sigma/\xi$ decreases, i.e., the optical obstacle becomes similar to a hard cylinder by stiffening its boundary.  

Our results are inconsistent with the $1/R$ dependence of $v_c$ that was predicted from analytic analyses on the stability of a superfluid flow~\cite{crescimanno,zwerger}. It is clearly seen that when $R/\xi(\approx \sigma/\xi)$ increases by a factor of 5, the $v_c/c_s$ decreases less than a factor of 2. Even without including the additional reduction effect due to the soft boundary of the obstacle, the critical velocity decreases much smaller than what would be expected from the $1/R$ dependence.

Finally, we want to recall a few aspects of the experimental condition that should be considered for the quantitative comparison of the measurement results to theoretical predictions for a 2D homogeneous case. First, the critical velocity was not measured for a steady flow condition but by sweeping a finite section of a trapped condensate. The measurement can be affected by the sweeping manner. For example, if vortex nucleation requires a finite time, which might be longer than the sweeping time near the critical velocity~\cite{berloff2,tsubota}, it would result in a systematic, upward shift of the measured value of $v_c$. Second, although we expect suppression of 3D vortex dynamics in a highly oblate condensate, it is an inevitable fact that the condensate has an inhomogeneous density distribution along the axial direction. Thus, one cannot completely ignore 3D responses of the condensate, in particular, at the moment of vortex nucleation. It might be necessary or sufficient to introduce an effective speed of sound that would be lower than the peak value $c_s$ of the condensate~\cite{huang}.

\section{Summary}  
  
We have presented the measurements of the critical velocity for vortex shedding in highly oblate Bose-Einstein condensates and investigated the soft boundary effect of the moving obstacle formed by a Gaussian potential. Our results are consistent with a picture based on the local Landau criterion and the measure value of $v_c/c_s$ with the stiffest obstacle is in good agreement with the theoretical predictions for a homogeneous 2D superflow past a cylindrical object. This work has established a reliable experimental method to measure the critical velocity of a trapped condensate and its intriguing extension is to investigate the temperature dependence of the critical velocity, which might provide a new setting to study the role of thermal atoms in vortex nucleation~\cite{griffin,gardiner,kasamatsu}.

\begin{acknowledgements} 

We thank Seji~Kang for experimental assistance. This work was supported by the National Research Foundation of Korea (Grant No. 2011-0017527).

\end{acknowledgements}

\end{document}